\documentstyle[multicol,prb,aps,epsf]{revtex}
\begin{document}
\title{Phonon spectral function for an interacting electron-phonon system}

\author{J.E. Han and  O. Gunnarsson} 
\address{Max-Planck-Institut f\"ur Festk\"orperforschung, 
D-70506 Stuttgart, Germany}

\date{\today}
\maketitle
%\pacs{71.20.Tx, 75.40C, 71.10.Fd}
\begin{abstract}
Using exact diagonalzation techniques, we study a model of interacting 
electrons and phonons. The spectral width of the phonons is found to be reduced
as the Coulomb interaction $U$ is increased. For a system with two modes 
per site, we find a transfer of coupling strength from the upper to the
lower mode. This transfer is reduced as $U$ is increased. These results
give a qualitative explanation of differences between Raman and
photoemission estimates of the electron-phonon coupling constants 
for A$_3$C$_{60}$ (A= K, Rb).
\end{abstract}
\begin{multicols}{2}
In a metallic system a phonon can decay into electron-hole 
pair excitations. This decay contributes to the width of the
phonon. It was pointed out by Allen that this additional broadening
can be used to estimate the electron-phonon coupling.\cite{Allen} 
The  width can be measured in neutron scattering or, for the 
orientationally disordered fullerenes, in Raman scattering.\cite{Jong}
Normally, the electron-phonon coupling is deduced by assuming
noninteracting electrons.\cite{Allen} The method is, however,
often applied to systems with strong correlation due to
the Coulomb interaction, such as the alkali-doped fullerenes.\cite{RMP}
In the alkali-doped fullerenes the electron-phonon interaction plays 
an important role, and accurate estimates of the coupling strength 
are essential. Almost all experimental estimates for these systems are 
based on Allen's formula, and the accuracy of this formula is 
therefore crucial. 

In strongly correlated systems the hopping is reduced and
the excitation of electron-hole pairs may be more 
difficult. For instance if the correlation is so strong that the 
system has a metal-insulator transition, the decay into electron-hole
pair excitations is completely suppressed. One aim of this paper is 
therefore to study how the estimate of the electron-phonon coupling 
is influenced if the electron-electron interaction is taken into 
account.

In metals          a phonon can decay into a (virtual) electron-hole 
pair excitations which can then decay into a different phonon. In this 
way there is a coupling between different phonon modes of the same
symmetry, leading to 
new modes which are linear combinations of the old ones. These 
new modes can have  quite different coupling strengths than the 
old modes. A second aim of this paper is to study how the coupling 
strength is transferred between the modes due to the coupling
via electron-hole pair excitations.

The electron-phonon coupling has been studied extensively for 
the alkali-doped fullerenes. In particular, there have been
a study based on neutron scattering,\cite{Prassides} and
several studies based on Raman scattering.\cite{Winter,Lannin}
The high resolution studies of Winter and Kuzmany\cite{Winter}
show a very strong coupling to a few of the low-lying modes,
but almost no coupling to the high-lying modes. An alternative 
approach is based on photoemission from free negatively charged
C$_{60}^-$ molecules. By studying the weight of 
vibration satellites, it is possible to deduce the electron-phonon
coupling.\cite{PES} This results in rather different electron-phonon
coupling constants. Although the main coupling was to the 
low-lying modes, there was also a substantial coupling to the two 
highest H$_g$ modes. The total coupling strength was also
larger than deduced from Raman scattering.  
 
In this paper we study a simple model with electron-phonon
and electron-electron interactions. We consider a finite
cluster with a nondegenerate electronic level and a nondegenerate
phonon on each site. This model is solved by using exact 
diagonalization. We find that the Coulomb interaction reduces
the phonon width, and that the use of Allen's formula\cite{Allen}
therefore leads to an underestimate of the electron-phonon 
coupling constants in Raman scattering experiments. Furthermore we 
find that due to the indirect interaction of different phonon modes 
via electron-hole pair excitations in metallic systems, there is a 
transfer of coupling strength to the low-lying modes which is not 
present for a free molecule. Since the Raman measurements
of the electron-phonon coupling are performed for a solid, but the 
photoemission estimate is for a free molecule, the weight transfer 
is present in the Raman but not in the photoemission estimate.  
These observations are consistent  with differences between 
the coupling constants deduced from Raman scattering and photoemission. 

We consider a model with $N_{\rm mode}$ nondegenerate phonons per 
site and with electrons without orbital degeneracy. The 
Hamiltonian is
\begin{eqnarray}\label{eq:1}
H&&=\sum_{i\nu}\omega_{\nu}b^{\dagger}_{i\nu}b_{i\nu}+\sum_{i\sigma} 
\lbrack \varepsilon_0 +\sum_{\nu}g_{\nu}           
(b_{i\nu}+ b_{i\nu}^{\dagger})\rbrack n_{i\sigma} \\ \nonumber
&&+U\sum_i n_{i\uparrow}n_{i\downarrow}+
\sum_{ij}t_{ij}c_{i\sigma}^{\dagger}c_{j\sigma},
\end{eqnarray}
where $i$ labels the $N_{\rm site}$ sites, $c_{i\sigma}$ and 
$b_{i\nu}$ annihilate an electron with spin $\sigma$ and a 
phonon with the label $\nu$, respectively, on site $i$ and
$n_{i\sigma}=c^{\dagger}_{i\sigma}c_{i\sigma}$ is an  occupation 
number operator. The energy of the phonon $\nu$ is $\omega_{\nu}$ and
its coupling to the electrons is is $g_{\nu}$.  The corresponding 
dimensionless electron-phonon coupling is given by 
$\lambda_{\nu}=2 g_{\nu}^2N(0)/\omega_{ph}$, where $N(0)$
is the density of states per spin. The energy of the electronic level
is $\epsilon_0$. Two electrons 
on the same site have a Coulomb repulsion $U$. The hopping between 
the sites is described by matrix elements $t_{ij}$.  
A Hamiltonian like (\ref{eq:1}) with $t_{ij}\equiv t$ for the 
nearest neighbor hopping  has a high symmetry and a correspondingly large 
degeneracy. Since we use exact diagonalization to solve the model, 
we have to limit the number of sites to a small number (4-6). 
The resulting one-particle states are then very sparse in energy. Therefore  
we lower the symmetry by choosing each $t_{ij}$ randomly within
some interval, which leads to a denser energy spectrum. The model 
is solved and the result is then averaged over different sets of 
$\lbrace t_{ij} \rbrace$.  The strength of the hopping is measured 
by the one-particle width $W$ of the electronic band (for $g=0$ and $U=0$).    
In this model we for simplicity consider nondegenerate (A$_g$) phonons
and electrons. In, for instance, A$_3$C$_{60}$ (A= K, Rb) the phonons 
are five-fold degenerate H$_g$ phonons and the electron states have 
a three-fold
orbital degeneracy. The new physics which may be introduced by these
degeneracies, e.g., the Jahn-Teller effect, is not considered here.

We consider a half-filled system, i.e., $N_{\rm site}$ electrons. 
With $N_{\rm site}=6$ there are then 400 different electronic configurations. 
To obtain a finite size Hilbert space, we limit the maximum number 
of phonons per modes to $N_{\rm phon}$. The number of phonon states
is then $(N_{\rm phon}+1)^{N_{\rm site}}$ for the case of one mode
per site. For instance, with $N_{\rm site}=6$ and $N_{\rm phon}=3$, 
the total Hilbert space has the dimension 
$1.6384\cdot 10^6$. Such a problem can be solved using exact diagonalization,
i.e., the ground-state is expressed as a linear combination of all possible 
basis states in the Hilbert space. The lowest eigenfunction of the 
corresponding Hamiltonian matrix is then found using the Lanczos
method. We further calculate the phonon Green's function
\begin{equation}\label{eq:2}
D^{\nu}_{ij}(t)=-i\langle 0|T\lbrace \phi_{i\nu}(t) \phi_{j\nu}
(0)\rbrace |0\rangle, \end{equation}
where $|0\rangle$ is the ground state, $\phi_{i\nu}(t)=b_{i\nu}(t)
+b_{i\nu}^{\dagger}(t)$
is the phonon field operator in the interaction representation
and $T$ is the time-ordering operator. Calculation of the Fourier transform
gives $D^{\nu}_{ij}(\omega)$. We then define a spectral function as       
\begin{equation}\label{eq:2a}
A^{\nu}_{ii}(\omega)={1\over \pi}|{\rm Im} D^{\nu}_{ii}(\omega)|,
\end{equation}
and study the average $\rho_{ph}(\omega)=\sum_{\nu i} A_{ii}i^{\nu}(\omega)
/N_{site}$.
Due to the finite size of the system, the spectrum is discrete.
We therefore introduce a Lorentzian broadening with the FWHM (full
width at half maximum) 0.01 eV.

\begin{figure}
\centerline{\epsfxsize=3.3in \epsffile{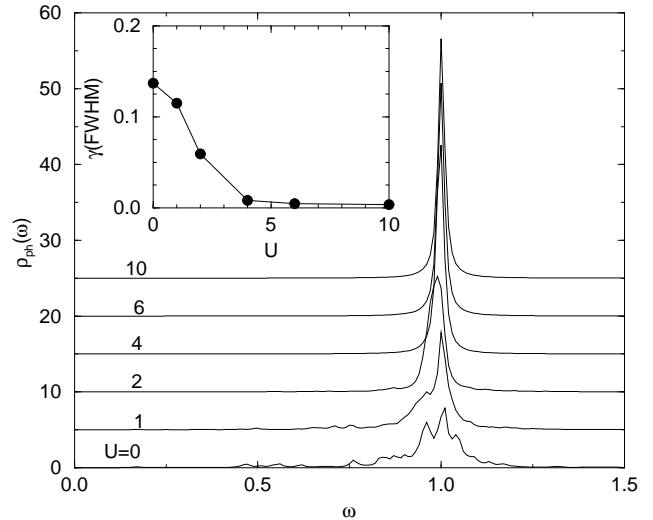}}
\caption[]{The phonon spectral function for different values of the  
Coulomb interaction $U$. The insert shows the width $\gamma$ of the
spectral function as a function of $U$. The figure illustrates the 
narrowing of the spectral function as $U$ is increased. The parameters are
$\omega_1=1$, $\lambda=0.073$ and $W=2.5$ and the system has six sites.}
\label{fig:1}
\end{figure}

Fig. \ref{fig:1} shows the phonon spectral function $A(\omega)$ 
for different values of $U$.
Due to the small size of the system, the width of the spectrum should 
not necessarily be expected to agree with Allen's formula even for $U=0$.
Nevertheless, the result of Allen's formula $\gamma_{\rm Allen}=0.19$,
is comparable to the width found for $U=0$.  
The figure illustrates how 
the spectral function becomes narrower with increasing $U$. This is 
further illustrated by the inset, which shows the width 
of the spectrum, calculated as the mean square deviation of the 
spectral function. The figure illustrates that one underestimates 
the electron-phonon coupling if Allen's formula is used to extract 
the coupling for a system with a finite $U$. 
For systems like $A_3$C$_{60}$ (A= K, Rb), where the Coulomb interaction
is believed to play an important role, the width of the phonons 
may then be substantially reduced. The use of Allen's formula
would then correspondingly underestimate the electron-phonon coupling.

We next discuss the case when there are two phonon modes per site,
which have the unperturbed energies $\omega_1$ and $\omega_2$.
First we calculate the lowest order phonon self-energy. This
involves evaluating  a ``bubble'' diagram. The self-energy
can be written as
\begin{equation}\label{eq:3}
\Pi_{\nu,\nu^{'}}({\bf q},\omega)\sim g_{\nu}g_{\nu^{'}}f(\omega),
\end{equation}
where $f(\omega)$ depends on the precise band structure.          
 We consider contributions to the self-energy which are
both diagonal and non-diagonal in the index $\nu$. The non-diagonal
contribution corresponds to a phonon $\nu$ decaying into an electron-hole
pair followed by this electron-hole pair decaying into a phonon 
$\nu^{'}$. The non-interacting phonon Green's function is
\begin{equation}\label{eq:4}
D^0_{\nu,\nu^{'}}(\omega)=2\omega_{\nu}/(\omega^2-\omega_{\nu}^2)
\delta_{\nu,\nu^{'}}. 
\end{equation}
The interacting phonon Green's function is then given by
\begin{eqnarray}\label{eq:5}
D^{-1}(\omega)&&=\lbrack D^0(\omega)\rbrack^{-1}-\Pi(\omega)= \\ 
&&\left[ \begin{array}{cc} {\omega^2-\omega_1^2\over 2\omega_1}-
g_1^2f(\omega)&-g_1g_2f(\omega)\\-g_1g_2f(\omega)&
{\omega^2-\omega_2^2\over 2\omega_2}-g_2^2f(\omega)\end{array}\right]
\nonumber
\end{eqnarray}
The modes of the coupled system are obtained by looking for zeros
of the determinant of the matrix in Eq. ~(\ref{eq:5}). For the 
lowest mode, the corresponding eigenvector consists of a
bonding linear combination of the two unperturbed modes. As a result
the coupling to the electrons is increased for these modes, due to
 constructive interference between the couplings for the two 
unperturbed modes. In the same way the coupling is reduced for
the higher mode. For instance, we can look for the width of 
the lowest mode in the limit when $\omega_2\gg \omega_1$ and 
when the electron-phonon coupling is weak. We then find that the width
of the lowest mode is increased by a factor of
\begin{equation}\label{eq:6}
(1+c\lambda_2),
\end{equation}
and the width of the highest mode is reduced by a factor
\begin{equation}\label{eq:7}
(1-c\lambda_2({\omega_1\over \omega2})^2),
\end{equation}
where $c$ is somewhat larger than unity ($c\sim 3$) and depends on 
the shape of the band.

\begin{figure}
\centerline{\epsfxsize=3.3in \epsffile{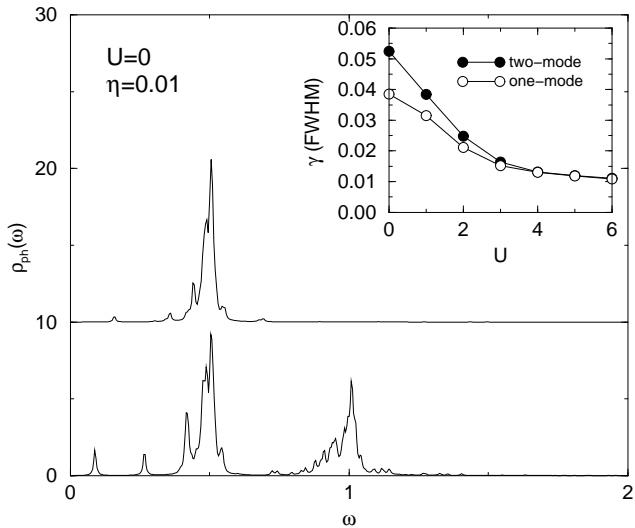}}
\caption[]{The phonon spectral function for a system with one 
phonon mode (upper part) and two phonon modes (lower part) 
per site for $U=0$. The insert
shows the width of the lower mode as a function of $U$ in the cases 
of one or two modes per site. Comparison of the widths in the two
cases, illustrates how the lower mode is broadened for 
small values of $U$ due to the interaction with the upper mode.
The parameters are $W=3.7$, $\omega_1=0.5$, $\omega_2=1$, 
$g_1=0.3$ and $g_2=0.4$ and the system has four sites.
All spectra have been given a Lorentzian broadening with the
FWHM=0.1  }
\label{fig:2}
\end{figure}

The result in Eq.~(\ref{eq:6}) is based on the lowest order phonon 
self-energy and it neglects the Coulomb repulsion completely. 
We therefore study the same problem using exact diagonalization.
Fig. \ref{fig:2} compares results for systems with one or
two modes per site. The discrete spectra have been broadened by
a Lorentzian with the FWHM=0.01. The main figure shows results for 
$U=0$, and it 
illustrates how the lower mode is broadened when the higher mode is 
switched on. For the parameters in Fig. \ref{fig:2} $\lambda_1
=0.043$ and $\lambda_2=0.085$, and the additional broadening of the lowest 
mode is of the order of magnitude predicted by Eq. (\ref{eq:6}). 
The insert shows the width of the lower mode as a function of $U$.
These results were obtained by fitting Lorentzians to the broadened 
spectra.  The width for very large values of $U$ is due to the 
broadening of the discrete spectrum that we have introduced.
As $U$ is increased, the width of the mode is reduced, as discussed
above. The figure further illustrates that the transfer of coupling
strength is reduced as $U$ is increased.  This is expected,
since the effects of hopping, and thereby the indirect coupling,
is reduced as $U$ is increased.

It would be interesting to repeat these calculations for systems
with degenerate phonons, e.g., to include the Jahn-Teller 
effect. This leads, however, to systems which are so large 
that they cannot easily be treated using exact diagonalization.
Within a Hartree calculation we find a similar transfer
of coupling strength to the lower modes also for Jahn-Teller phonons
and electrons with orbital degeneracy. The transfer is, however, 
reduced by the nonspherical parts of the Coulomb interaction, i.e., 
by the difference between the interaction for equal orbitals and  
unequal orbitals. This effect may also play a role when we go
beyond the Hartree approximation. 

Finally, we observe that in theoretical approaches which do not 
explicitly include the transfer of coupling strength between the modes, 
it is appropriate to include this transfer by using the corresponding 
coupling constants. On the other hand, in a treatment where this transfer 
is explicitly included, the transfer should not be contained in the 
coupling constants used in the model. 

To summarize, we have calculated the phonon spectral functions
for systems with interacting electrons and phonons. We find
that the Coulomb interaction between the electrons reduces the
width of the phonons caused by the phonon decay into electron-hole 
pairs. As a result, estimates of the electron-phonon coupling based 
on the phonon width underestimate this coupling unless the Coulomb 
interaction is taken into account. This is consistent with the   
observations that weaker couplings have been deduced from Raman
measurements than from
photoemission (PES) experiments. Furthermore, we find that there is
a transfer of coupling strength from the higher modes to the 
lower modes due to an indirect interaction via electron-hole
pairs. This may, at least partly, explain the difference 
in the distribution of coupling strength between Raman and
PES estimates, although it can probably not fully explain the
weak coupling to the two highest phonons seen in Raman spectroscopy. 
In this  work we have treated nondegenerate phonons. 
It would be interesting to extend the  work to degenerate, Jahn-Teller
phonons, since these are the important phonons in the alkali-doped 
Fullerenes. 

This work has been supported by the Max-Planck-Forschungspreis.

\end{multicols}
\end{document}